\documentclass[3p,times]{elsarticle}

\usepackage{ecrc}

\usepackage{color}

\usepackage{epstopdf}

\volume{00}

\firstpage{1}

\journalname{Nuclear Physics A}

\runauth{V. Loggins}

\jid{nupha}

\jnltitlelogo{Nuclear Physics A}

\usepackage{subcaption,graphicx}
\usepackage{caption}
\usepackage{subcaption}
\usepackage{amsmath,amssymb}

\begin{document}

\begin{frontmatter}



\title{Azimuthally differential pion femtoscopy in Pb-Pb collisions at 2.76 TeV with ALICE at the LHC}

\author{Vera Loggins (for the ALICE\fnref{col1} Collaboration)}
\fntext[col1] {A list of members of the ALICE Collaboration and acknowledgements can be found at the end of this issue.}
\address{Wayne State University, 42 W. Warren Ave, Detroit, MI 48202}




\begin{abstract}
Femtoscopy of non-central heavy ion collisions provides access to information on the geometry of the effective pion-emitting source. 
In particular, the source shape can be studied by measuring femtoscopic radii as a function of the pion emission angle relative to the collision symmetry planes.  
We present the results of azimuthally differential femtoscopy of Pb-Pb collisions {\color{black}at} $\sqrt{s_{\textup{NN}}}=2.76$~TeV at the LHC{\color{black}~relative to the second harmonic event plane}. 
%
We observe {\color{black}a} clear oscillation of the extracted radii as a function of  the emission angle. 
We find that $R_{side}$ and $R_{out}$ oscillate out of phase for all centralities and pion transverse momenta.
The relative amplitude of $R_{side}$ oscillation decreases in more central collisions, but remains positive, which indicates that the source remains out-of-plane extended qualitatively similar to {\color{black} what was} observed at RHIC energies. We compare our results to existing hydrodynamical and transport model calculations.

\end{abstract}


\end{frontmatter}


\color{black}

\section{Introduction}
\label{intro}

Two particle correlations at small relative momenta (commonly known as \emph{femtoscopy}) is an effective tool to probe the space and time characteristics of the particle emitting source in relativistic heavy ion collisions~\cite{ref1,ref2}.
The results presented here are obtained in the so-called Longitudinal CoMoving System (LCMS) in which the total pair momentum along the \textit{z} direction is zero, ($p_{1,z}=-p_{2,z}$).
In this system, the extracted freeze-out radii provide information on the system evolution in the following manner:  $R_{side}$ is mostly determined by the system geometrical size, $R_{out}$ is mostly determined by the system geometrical size and the emission duration, and $R_{long}$ is mostly determined by the total emission time.
~Dependence of the radii on the transverse momentum provides information on the system's collective radial expansion (flow)~\cite{ref4}.  
{\color{black}Flow is due to pressure gradients which are higher in the plane of reaction (in-plane) than perpendicular to it (out-of-plane). Due to flow, the freeze-out source shape may be less out-of-plane extended, and more in-plane extended~\cite{ref4}. In this analysis, we focus on the azimuthal dependence of the radii with respect to the reaction plane to get information on the shape of the source at freeze-out.}

\section{Data Analysis}
\label{Analysis}

The data sample used for this analysis is recorded by ALICE during the 2011 heavy-ion run, Pb-Pb collisions at $\sqrt{s_{NN}}=2.76$~TeV.  There are roughly 30 million {\color{black}min-bias and triggered (central and semi-central)} events used in this analysis.  The tracks are reconstructed using the Time Projection Chamber (TPC)~\cite{ref5}.  Both the TPC and the Time of Flight (TOF)~\cite{ref5}~ were used for {\color{black}pion} identification in the pseudorapidity range $|\eta|<0.8$.  The correlation function is defined as the ratio of signal and background {\color{black}relative momentum }distribution of two identical pions.
The signal distribution was formed using particles from the same event, whereas the background distribution was formed using particles from different events.  
The pair cuts {\color{black}have been applied} to reduce ``track splitting" (false pairs created at low relative momentum) and ``track merging" (when two tracks are reconstructed as one).  

\section{Azimuthally Differential Pion Femtoscopy Results}

Azimuthally differential femtoscopic analysis of pion production relative to the second harmonic event plane has been performed for Pb-Pb collisions at $\sqrt{s_{NN}}=2.76$~TeV.  Femtoscopic radii have been extracted from the fit to the correlation function using Bowler-Sinyukov fitting procedure~\cite{ref6}:
%
\begin{eqnarray}
 C({\vec q},\Delta\phi)=N[(1-\lambda)+\lambda K({\vec q})(1+G({\vec q},\Delta\phi)],
 \end{eqnarray}
where $G({\vec q},\Delta\phi)=e^{-q_{out}^{2}R_{out}^{2}(\Delta\phi)-q_{side}^{2}R_{side}^{2}(\Delta\phi)-q_{long}^{2}R_{long}^{2}(\Delta\phi)-q_{out}q_{side}R_{\textit{os}}^{2}(\Delta\phi) }$, \textit{N} is the normalization parameter,  $K({\vec q})$ is the Coulomb component, $\lambda$ is the fraction of pairs participating in the Bose-Einstein correlation, and $\Delta\phi=\varphi_{pair}-\Psi_{EP,2}$ is the relative pair angle with respect to the {\color{black}second harmonic event plane defined by the TPC tracks}.  The extracted radii as a function of $\Delta\phi$ are fitted by the following Fourier expansion:
 \begin{eqnarray} \label{Radii_osc}
 \begin{split}
 R^{2}_{\mu}=R^{2}_{\mu,0}+2 R^{2}_{\mu,2}\cos[2(\Delta\phi)]\;\;\;(\mu=out,side,long), \\
  R^{2}_{\mu}=2 R^{2}_{\mu,2}\sin[2(\Delta\phi)]\;\;\;(\mu=\textit{out-side}).
  \end{split}
  \label{radiiEQ}
 \end{eqnarray}
 
 \begin{figure}[htp]
\centering
  {\includegraphics[width=7.5cm]{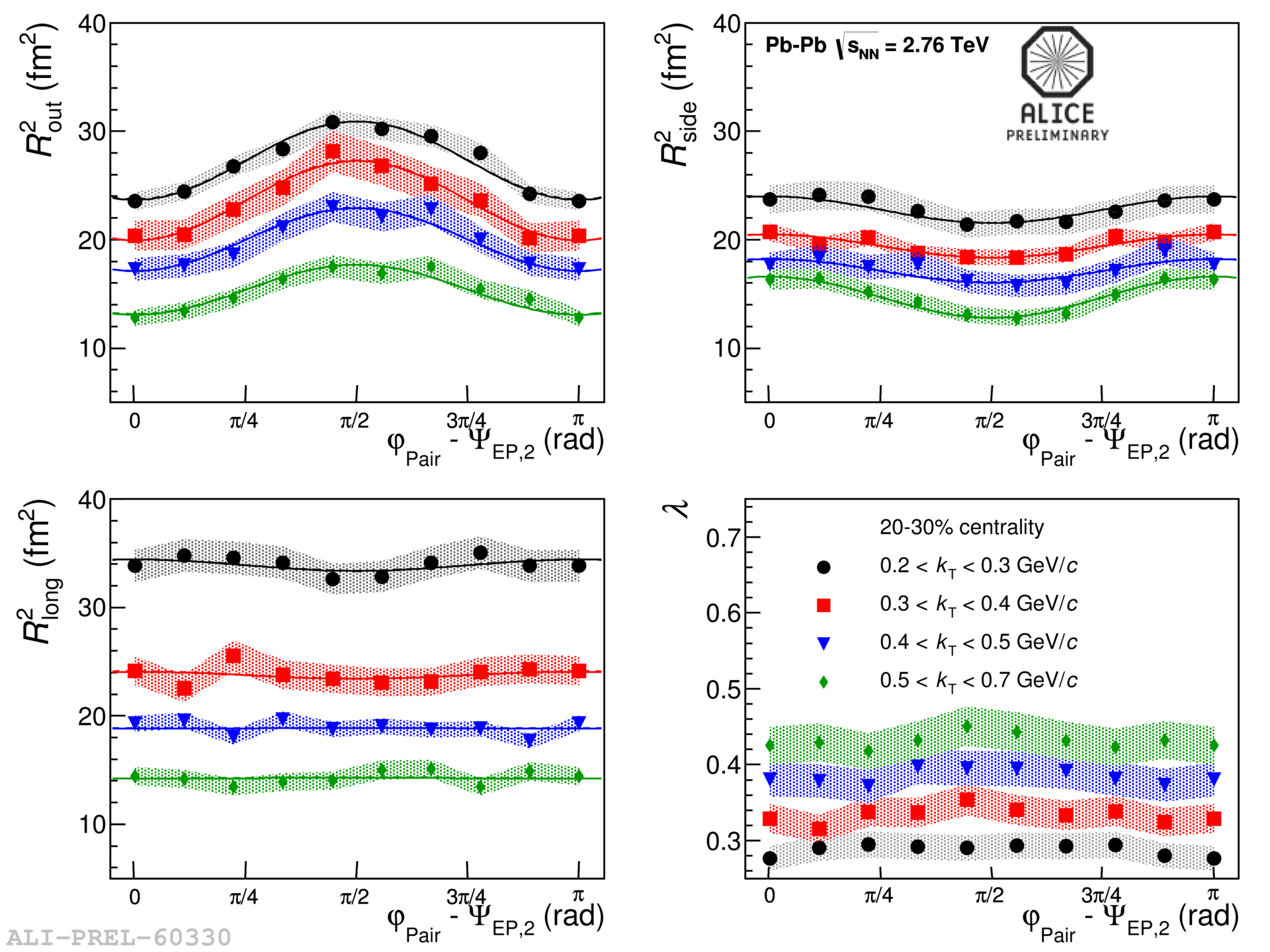}}\hspace{1em}%
  {\includegraphics[width=7.5cm]{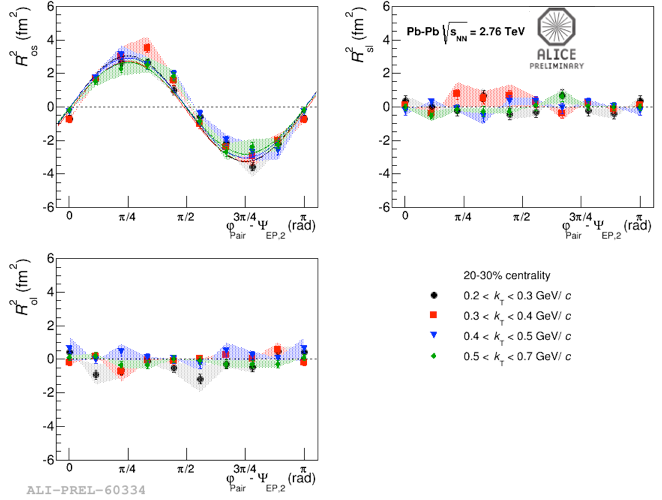}}
  \caption{
$R_{out}^{2}$, $R_{side}^{2}$, $R_{long}^{2}$, and $\lambda$ (left) and $R_{os}^{2}$, $R_{sl}^{2}$, and $R_{ol}^{2}$ (right) at 20-30\% centrality as a function of emission angle $\Delta\phi$ for different $k_{\textup{T}}$ intervals:~0.2-0.3, 0.3-0.4, 0.4-0.5, and 0.5-0.7 GeV/$c$.  Lines represent the fits to the radii using Eq.~\ref{radiiEQ}.  The statistical errors are shown by the error bars and systematic errors are indicated by shaded regions.}
\label{ktdep_radiii_cross}
\end{figure}

Figure~\ref{ktdep_radiii_cross} (left) presents the {\color{black}$R_{out}^2$,~$R_{side}^2$,~$R_{long}^2$} radii {\color{black}and $\lambda$} dependence on transverse momentum $k_{\textup{T}}$, ($k_{\textup{T}}=p_1+p_2/2$) {\color{black}at} 20-30\% centrality.~
Figure~\ref{ktdep_radiii_cross} (right) shows {\color{black}similar} dependence of  \textit{out-side}, $R_{os}$, \textit{side-long}, $R_{sl}$, and \textit{out-long}, $R_{ol}$, the cross-radii.  
%
The values of the radii $R_{out}^2$, $R_{side}^2$, and $R_{long}^2$ {\color{black}at} $0.2<k_{\textup{T}}<0.3~$GeV/$c$ are higher than the values of the radii obtained at {\color{black}a} high $k_{\textup{T}}$, $0.5<k_{\textup{T}}<0.7$~GeV/$c$, {\color{black}as expected due to radial flow}.  
The oscillations for $R_{out}^2$ and $R_{side}^2$ are out-of-phase {\color{black} at all measured $k_{\textup{T}}$ ranges.
%
~We compare results obtained at $\sqrt{s_{\textup{NN}}}=2.76$~TeV {\color{black} in} Pb-Pb collisions to results obtained at RHIC~\cite{ref8} {\color{black}in} Au-Au collisions at $\sqrt{s_{NN}}=200$~GeV {\color{black}in Figure~\ref{fig:alicestar1stktrangex}}.  The radii oscillations are similar between RHIC and LHC.~
{\color{black}The results contradict {\color{black}to AZHYDRO calculations}{\color{black}~\cite{ref10} that predicted}  a sign inversion in the oscillation amplitude for $R_{side}$ at low $k_{\textup{T}}$.

%
{\color{black}Figure~\ref{fig:3_1D_HYDRO} shows comparisons of $R_{side,0}^2$, (left), and $R_{side,2}^2$/$R_{side,0}^2$, (right), with {\color{black}the most recent (3+1D) }hydrodynamical calculations~\cite{ref9}.
As was shown in~\cite{ref4}, 2$R_{side,2}^2$/$R_{side,0}^2$ at small $k_{\textup{T}}$ can be used as an estimate for the {\color{black}final} freeze-out eccentricity.}
We compare the final source eccentricity with experiments at lower energies in Figure~\ref{fig:final_eccent}.
%
{\color{black}Final source eccentricity is lower at higher energies as expected due to longer evolution time~\cite{ref11}.  HYDRO calculations~\cite{ref12} predict much stronger energy dependence than observed experimentally.  UrQMD model~\cite{ref7} describes the energy dependence of the final source eccentricity rather well, but it fails to describe $R_{side,0}$ and $R_{side,2}$ separately (not shown).}

\begin{figure}
\begin{center}
\includegraphics*[width=9.1cm]{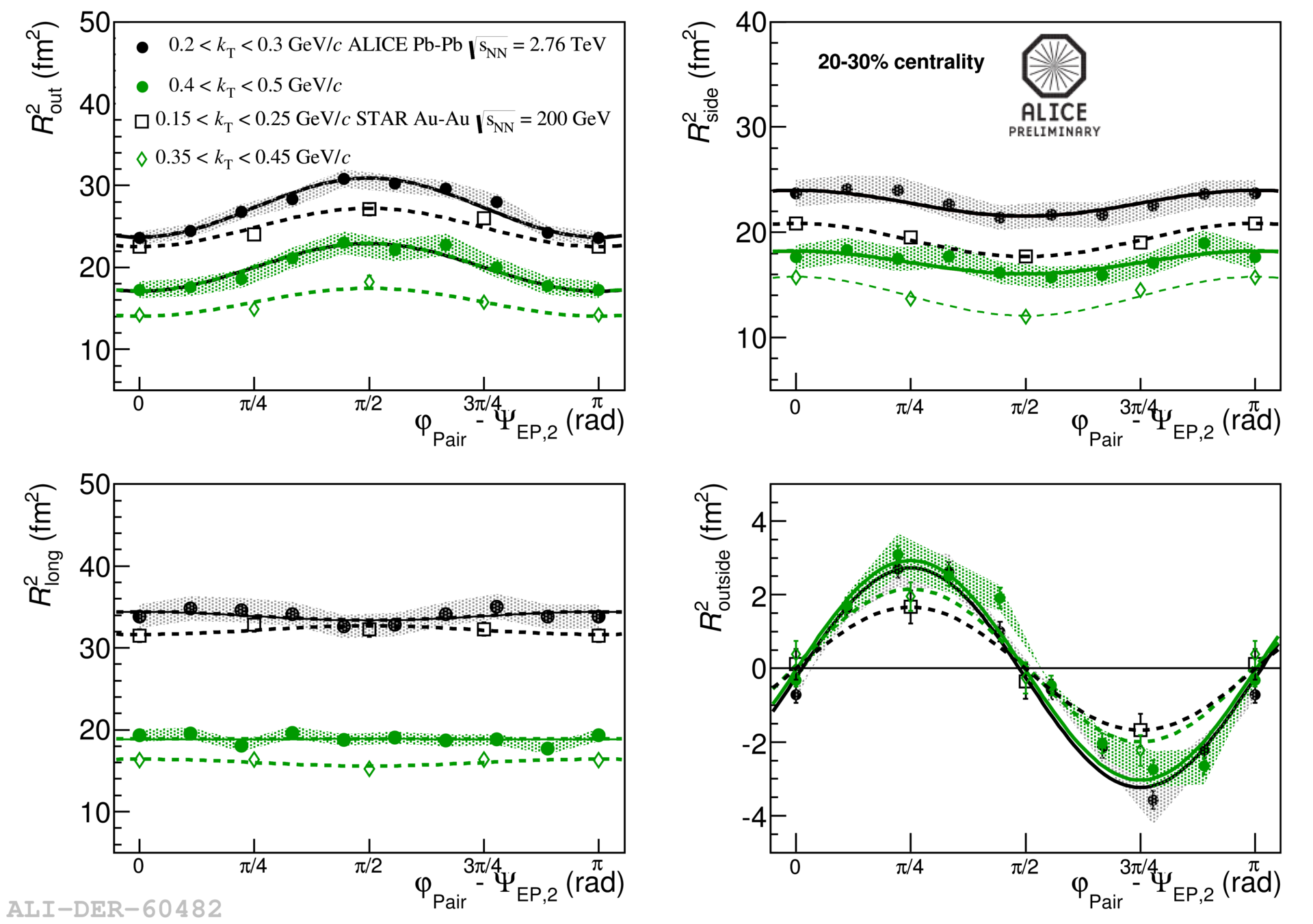}
\caption{
Centrality comparison of radii vs. pair emission angle at ALICE $k_{\textup{T}}$ of 0.2-0.3 GeV/$c$, 0.4-0.5
GeV/$c$, and STAR $k_T$ ranges 0.15-0.25 GeV/$c$, and 0.35-0.45 GeV/$c$~\cite{ref8} for 20-30\% centrality. The statistical errors are shown by the error bars and the systematic errors are indicated by shaded regions.
}
\label{fig:alicestar1stktrangex}
\end{center}
\end{figure}


 \begin{figure}[htp]
\centering
  {\includegraphics[width=7.0cm]{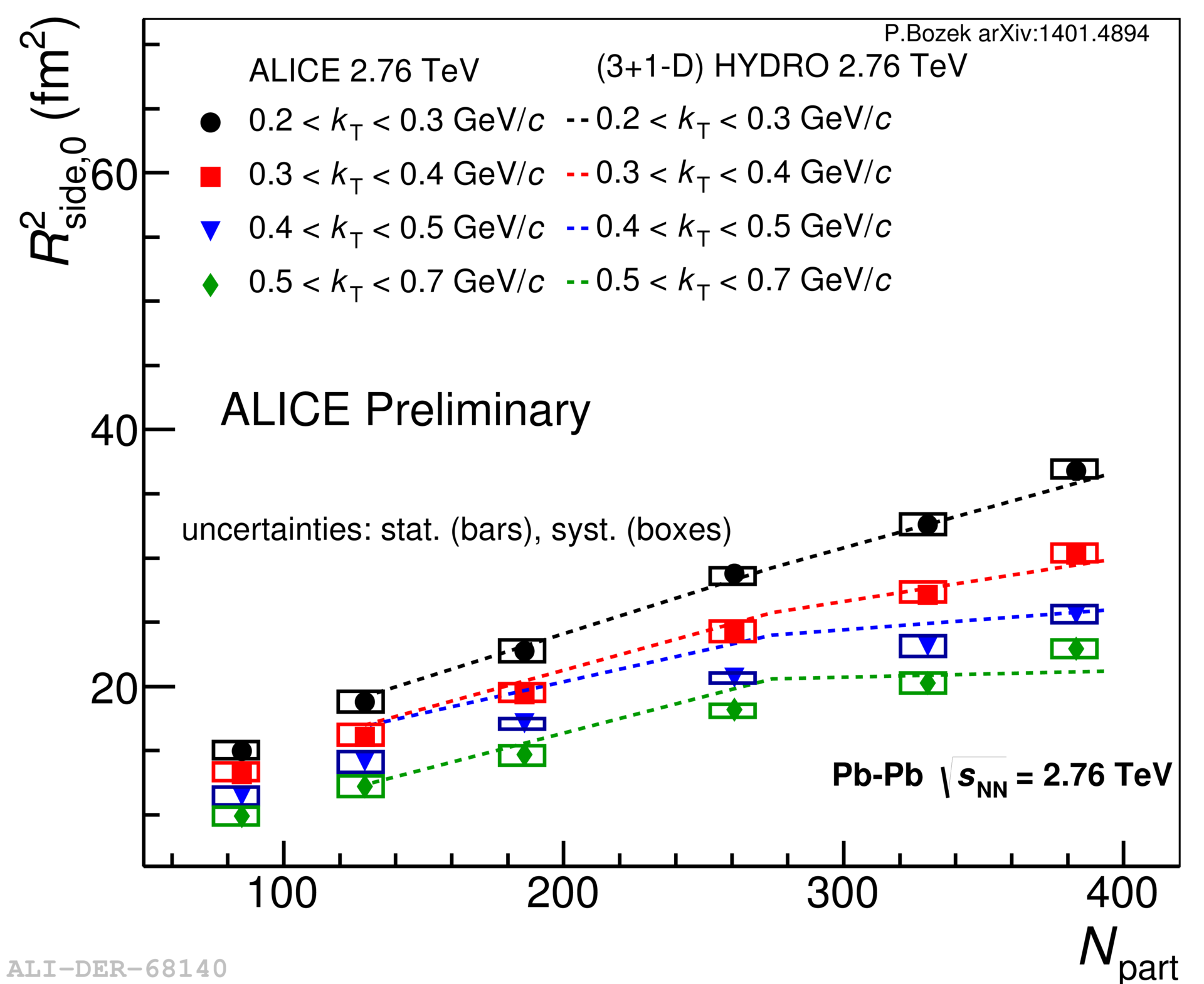}}\hspace{1em}%
  {\includegraphics[width=7.0cm]{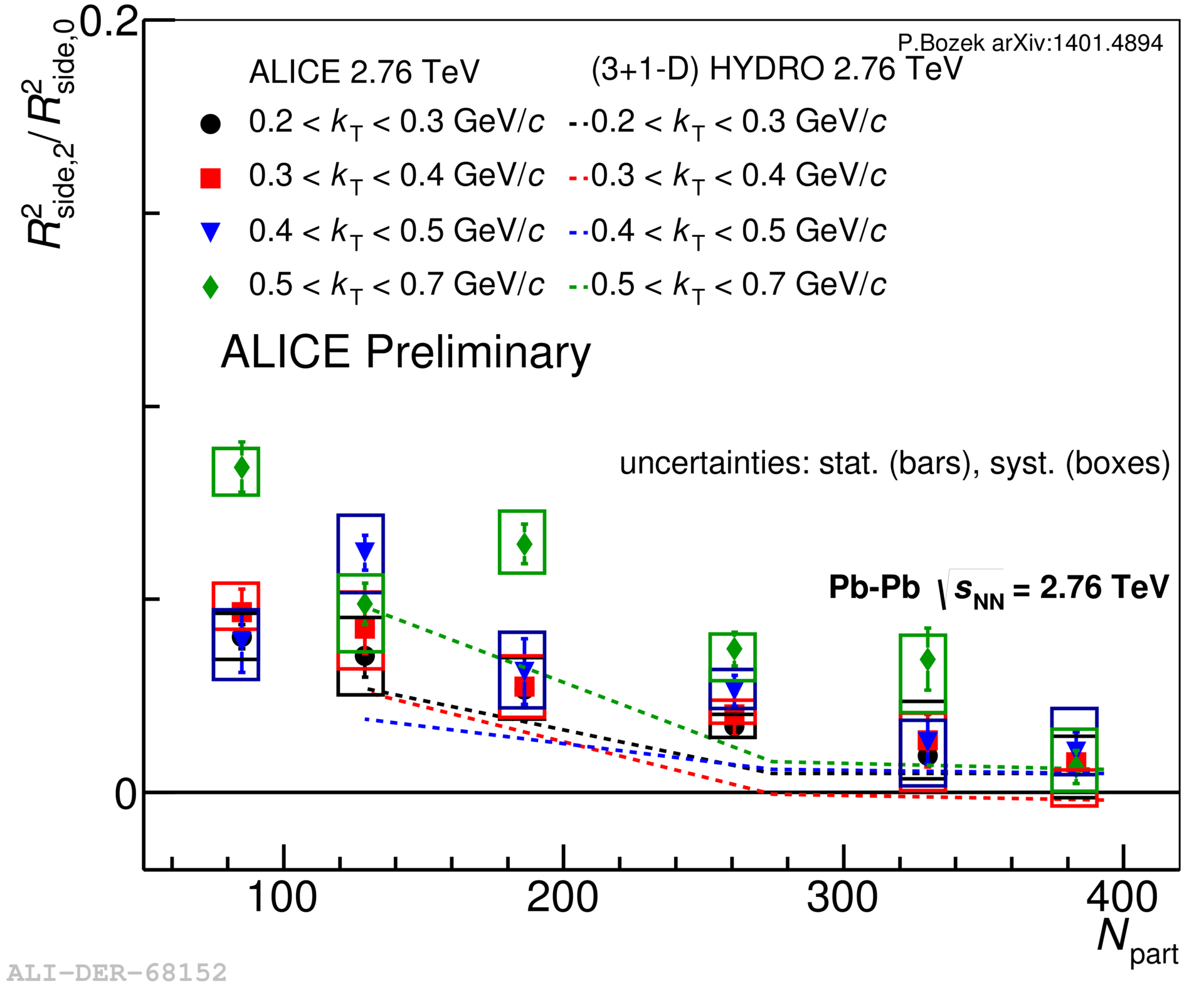}}
  \caption{
(3+1D) HYDRO model by P. Bozek~\cite{ref9} compared to $R_{side,0}^2$ (fm$^{2}$) (left) and $R_{side,2}^2$/$R_{side,0}^2$ (right). The statistical errors are shown by the error bars and the systematics are indicated by boxes. }
\label{fig:3_1D_HYDRO}
\end{figure}

 \begin{figure}
\begin{center}
\includegraphics*[width=9.8cm]{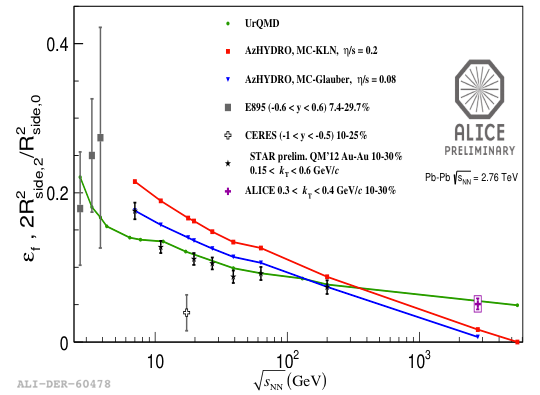}
\caption{
$\sqrt{s_{\textup{NN}}}$ dependence of the final spatial eccentricity. The statistical ({\color{black}systematic}) errors are shown by the error bars (boxes).}
\label{fig:final_eccent}
\end{center}
\end{figure}

\newpage

\color{black}{
\section{Summary}
We have performed an analysis of azimuthally differential pion femtoscopy relative to the second-order event plane for Pb-Pb collisions at $\sqrt{s_{\textup{NN}}}$=2.76~TeV.~
{\color{black}The radii oscillations as a function of two particle emission angle relative to the second order harmonic plane are found very similar at LHC to those at RHIC.~
The relative amplitude of oscillation, $R_{side,2}^2/R_{side,0}^2$ is found to decrease with collision energy reflecting lower source eccentricity in the freeze-out stage.}
 ~We obtain a positive value for $\epsilon_{f}$ indicating no change in the source geometry at freeze-out.  In the future, these observations will be studied further with additional measurements at higher energies.
}
%
%
%
%
%
%

\end{document}